\documentclass[10pt,conference,letterpaper]{IEEEtran}
\IEEEoverridecommandlockouts
\usepackage{cite}

\usepackage{amsmath,amssymb,amsfonts}
\usepackage{algorithmic}
\usepackage{graphicx}
\usepackage{textcomp}
\usepackage{xcolor}
\usepackage[ruled,linesnumbered]{algorithm2e} 
\usepackage[nolist]{acronym}
\usepackage{soul}
\usepackage{marginnote}
\usepackage{url}
\usepackage[inline]{enumitem}
\usepackage{tabularx}
\usepackage{amsthm}
\usepackage{tikz}

\newcommand{\myssec}[2]{%
  \subsection{#1}%
  \label{sec:#2}%
}
\newcommand{\rsec}[1]{%
  Sec.~\ref{sec:#1}%
}

\newcommand{\added}[1]{%
  {\color{blue}#1%
  }%
}
\renewcommand{\added}[1]{#1}
\newcommand{\addedb}[1]{%
  {\color{blue}#1%
  }%
}
\renewcommand{\addedb}[1]{#1}
\newcommand{\removed}[1]{%
  {\color{red}#1%
  }%
}
\renewcommand{\removed}[1]{}

\newcommand{\myfigeps}[3][width=\columnwidth]{%
\begin{figure}[tbp]%
\centering%
\includegraphics[#1]{figures/#2}%
\caption{#3}%
\label{fig:#2}%
\end{figure}%
}

\newcommand{\myfigfulleps}[3][width=\textwidth]{%
\begin{figure*}[tb]%
\centering%
\includegraphics[#1]{figures/#2}%
\caption{#3}%
\label{fig:#2}%
\end{figure*}%
}

\newcommand{\rfig}[1]{Fig.~\ref{fig:#1}}

\newenvironment{myinlinelist}%
{%
\begin{enumerate*}[label=(\roman*)]%
}%
{%
\end{enumerate*}%
}

\newenvironment{myitemlist}%
{%
\begin{itemize}[parsep=0em,leftmargin=*,label={--}]%
}%
{%
\end{itemize}%
}

\newenvironment{myenumlist}%
{%
\begin{enumerate}[parsep=0em,leftmargin=*,label=\arabic*.]%
}%
{%
\end{enumerate}%
}

\begin{document}

\title{
  Quality of Service in Quantum Networks

}

\author{
\IEEEauthorblockN{Claudio Cicconetti}
\IEEEauthorblockA{\textit{IIT, CNR} --
Pisa, Italy \\
c.cicconetti@iit.cnr.it}
\and
\IEEEauthorblockN{Marco Conti}
\IEEEauthorblockA{\textit{IIT, CNR} --
Pisa, Italy \\
m.conti@iit.cnr.it}
\and
\IEEEauthorblockN{Andrea Passarella}
\IEEEauthorblockA{\textit{IIT, CNR} --
Pisa, Italy \\
a.passarella@iit.cnr.it}
}

\author{Claudio~Cicconetti,
        Marco~Conti,
        and Andrea~Passarella%
\IEEEcompsocitemizethanks{\IEEEcompsocthanksitem All the authors are with the Institute of Informatics and Telematics (IIT) of the National Research Council (CNR), Pisa, Italy.}%
}

\IEEEtitleabstractindextext{%
\begin{abstract}
  In the coming years, quantum networks will allow quantum applications to thrive thanks to the new opportunities offered by end-to-end entanglement of qubits on remote hosts via quantum repeaters.
On a geographical scale, this will lead to the dawn of the Quantum Internet.
While a full-blown deployment is yet to come, the research community is already working on a variety of individual enabling technologies and solutions.
In this paper, with the guidance of extensive simulations, we take a broader view and investigate the problems of Quality of Service (QoS) and provisioning in the context of quantum networks, which are very different from their counterparts in classical data networks due to some of their fundamental properties.
Our work leads the way towards a new class of studies that will allow the research community to better understand the challenges of quantum networks and their potential commercial exploitation.
\end{abstract}

}

\maketitle

\begin{tikzpicture}[remember picture,overlay]
\node[anchor=south,yshift=10pt] at (current page.south) {\fbox{\parbox{\dimexpr\textwidth-\fboxsep-\fboxrule\relax}{
  \footnotesize{
\textcopyright 2022 IEEE.  Personal use of this material is permitted.  Permission from IEEE must be obtained for all other uses, in any current or future media, including reprinting/republishing this material for advertising or promotional purposes, creating new collective works, for resale or redistribution to servers or lists, or reuse of any copyrighted component of this work in other works.
  }
}}};
\end{tikzpicture}

\IEEEdisplaynontitleabstractindextext

\IEEEpeerreviewmaketitle

\bibliographystyle{abbrv}


%
  \section{Introduction}%
  \label{sec:introduction}%
Decades after it was theorized, \ac{QC} has finally emerged as a viable solution with the potential to revolutionize science, industry, and society~\cite{sevilla_forecasting_2020}.
Today, \ac{QC} is being offered as-a-service by very few large companies, led by IBM, but we can expect that in the future quantum computers will become more affordable and undergo mass adoption.
%
%
\removed{
When this happens, their interconnection via a global quantum network, the \textbf{Quantum Internet}, will unlock further possibilities, such as \textit{blind \ac{QC}} and \textit{distributed \ac{QC}}~\cite{wehner_quantum_2018}.
Meanwhile, in a shorter term, the Quantum Internet will enable high-impact applications not related to computing, e.g., \textit{quantum sensing}~\cite{pirandola_advances_2018} and \textit{\ac{QKD}}~\cite{pedone_toward_2021}.
}
\added{
When this happens, their interconnection via a global quantum network, the \textbf{Quantum Internet}, will unlock further possibilities, as envisioned by the \ac{IRTF} \ac{QIRG}, which is leading the standardization efforts towards the definition of a widely-adopted architecture and the identification of use cases of practical interest.
}

The basic unit of computation in \ac{QC} is the qubit, which unlike a classical bit can be in a \textit{superposition} of independent states and can be \textit{entangled} with other qubits.
Qubits can never be copied, but only teleported from one system to another through a process that unavoidably destroys the source state as soon as it is re-created on the destination.
Teleporting is the first step towards the realization of end-to-end entanglement.
The ultimate goal of a quantum network is: enabling a qubit on a local host to be entangled with another on a remote host, with the two hosts interconnected by \textit{quantum repeaters}.
\removed{In fact, qubits degrade during the transfer from source to destination and it is not possible  to regenerate/amplify them along the way (like with classical bits), so the only viable solution is to use intermediate devices with fresh qubits at each hop.}
\added{%
A vibrant research community has been working towards the realization of quantum repeaters since their inception more than 20 years ago~\cite{briegel_quantum_1998-1}, but the task has proved very challenging.
Only recently, technology developments have made it possible to obtain encouraging results using both optical fibers (\cite{pompili_realization_2021}) and free-space via satellite links (\cite{yin_satellite-based_2017}).
In particular, the latter is envisioned as a practical solution to interconnect quantum computers on geographical distances~\cite{de_parny_satellite-based_2022}.
}

Even though quantum repeaters are not yet commercially available, in the \ac{QC} community there is a widely accepted roadmap that foresees their gradual development through a series of upcoming \textit{generations}~\cite{munro_inside_2015}.
\removed{%
The definition of the architecture of the Quantum Internet and its protocols will have to adapt to the technology evolution, as well as to the future needs of the upcoming applications and services that will emerge in the market.
}
\added{%
However, \ac{QoS} in quantum networks needs significantly different approaches with respect to classical networks, as key enabling concepts (such as throughput and delay) have to be re-defined, while others (such as fidelity, introduced in \rsec{background:basics}) are completely new.}
\removed{%
One of the most investigated aspects is \textit{quantum routing}: the selection of the best path to establish end-to-end entanglement of qubits between two hosts (data plane) coupled with a procedure to implement it (control plane).}
\added{%
Furthermore, network provisioning is greatly affected by the fragility of qubits and the inherent impossibility to copy them.
Despite the pivotal role these aspects will play to the commercial success of the first Quantum Internet deployments, they have received very little attention so far.
To fill this gap, in this work we tackle \ac{QoS} and network provisioning in quantum networks in a wide range of conditions, considering the costs induced by quantum repeaters \addedb{in terms of capacity or fidelity.}}
Our analysis is carried out with the help of many simulated scenarios, playing the role of surrogates for real-world deployments in their absence.




The rest of this paper is structured as follows.
In \rsec{background} we introduce our system model and assumptions, followed by a review of the essential state of the art, in \rsec{soa}.
In \rsec{contribution} we analyze the key aspects of \ac{QoS} provisioning in quantum networks as emerging from a selection of broad results obtained with simulation.
A summary of the findings is presented in \rsec{conclusions}, accompanied by a discussion on important open research directions.
  \section{Background}%
  \label{sec:background}%
  In this section we briefly survey some fundamental aspects of quantum computing and communication (\rsec{background:basics}), and then we introduce the system model of quantum networks used in this work (\rsec{background:model}).

\myssec{Basics of quantum computing and networking}{background:basics}

The unit of computation of \ac{QC} is the \textit{qubit}, which represents any superposition of two different states (e.g., `0' and `1') with given (complex) probabilities depending on its initial preparation and subsequent operations.
When the qubit is measured, its state collapses to a well-defined state, which is the outcome of the measurement, that is a classical bit `0' or `1', and it is not possible to restore the stochastic state it had before this operation.
A qubit can be \textit{entangled} with another, which means that they share an intrinsic correlation: the state of one qubit cannot be described independently from that of the other one, even if they become separated by an arbitrarily large distance.
%
\removed{In general, entanglement is the basic ingredient for the realization of most \ac{QC} applications.}
%
Indeed, the goal of a quantum network (or the Quantum Internet, at a global scale) is to enable the entanglement of qubits in different locations by means of devices called \textit{quantum repeaters}.
\added{The latter are quantum computers, with a limited set of operations, which can transfer the state of a qubit across different media:}
\begin{myitemlist}
    \item Quantum computers perform local operations on so-called \textit{matter qubits} made of circuits of superconducting material\removed{that encode the state as two quantized energy levels operating at exceedingly low temperatures, only slightly above absolute zero}.
    Matter qubits are stored in quantum memories, which are interconnected by logic that makes them interact via quantum gates for the execution of local algorithms.
    \item Quantum communication devices physically transfer from one system to another \textit{flying qubits}, which most often are made of single photons encoding the state in the polarization (horizontal vs.\ vertical), which can be carried over fiber optic cables at normal temperatures, even though some cooling is required especially on the receiving side to reduce noise effects~\cite{tomm_bright_2021}.
\end{myitemlist}

\myfigeps{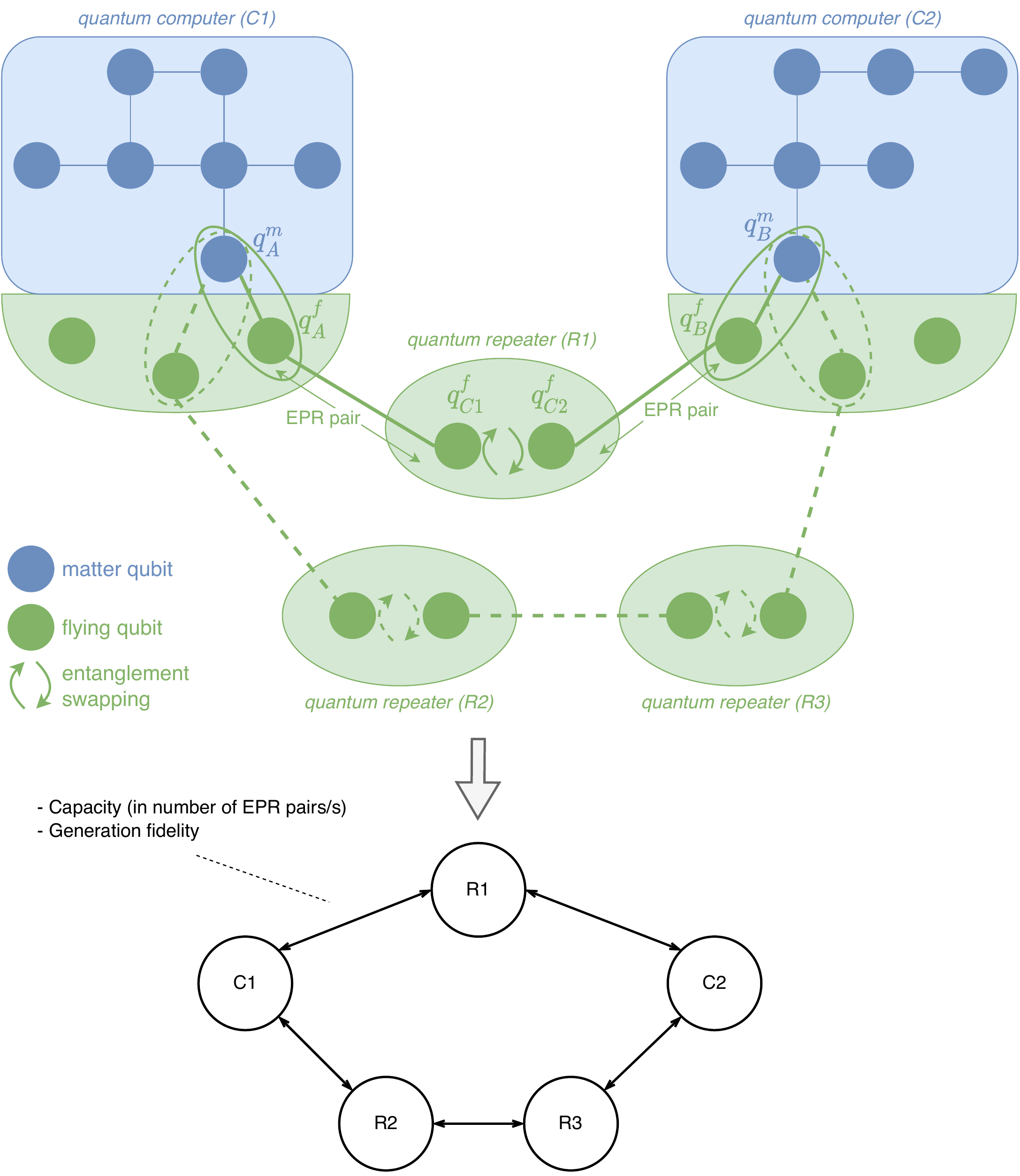}{%
\underline{Top}: Example of a minimal quantum network interconnecting two hosts $C1$ and $C2$: the quantum memories are in blue, like their quantum qubits, and the links between them are a high-level representation of the specific chip physical topology; the quantum communication devices (and the flying qubits) are in green, with links representing the medium (e.g., fiber optic) interconnecting to a quantum memory or a neighboring quantum repeater. There are two possible paths to establish end-to-end entanglement between the remote matter qubits $q^m_A$ and $q^m_B$: via $R1$ or via $R2$--$R3$.
\underline{Bottom}: Model corresponding to the above quantum network, with each link characterized by the maximum number of EPR pairs/s that can be generated and used for the entanglement (capacity), and the generation fidelity.}

Overall, a quantum network looks like the example in \rfig{basics-1} (top part), with two hosts ($C1$ and $C2$) interconnected by quantum repeaters ($R1$--$R3$).
Let us see what happens when a customer requires end-to-end entanglement between qubits on $C1$ and $C2$, say matter qubits $q^m_A$ and $q^m_B$, assuming entanglement path $C1$--$R1$--$C2$:
%
\begin{myenumlist}
    \item Two EPR pairs $\langle q^f_A,q^f_{C1} \rangle$ and $\langle q^f_B,q^f_{C2}\rangle$ are shared between the repeater $R1$ and the two end hosts $C1$ and $C2$, respectively.
    An \textit{EPR pair} is a system of two qubits whose states are maximally entangled, that is, informally, they are as strongly correlated as possible, the name being a tribute to Einstein, Podolsky, and Rosen.

    \item $R1$ performs \textit{entanglement swapping} on $q^f_{C1}$ and $q^f_{C2}$, which is a local operation, involving a measurement, with the effect of entangling the other qubits in the two EPR pairs, i.e., $q^f_A$ and $q^f_B$.
    The process produces classical bits, which are transmitted to $C1$ (or equivalently $C2$).

    \item $C1$ entangles locally its matter qubit $q^m_A$ with $q^f_A$.
    The viability of this step has been demonstrated in laboratory conditions, e.g., in \cite{pompili_realization_2021} using diamond \ac{NV} centers.
    The same happens on $C2$ for $q^m_B$ and $q^f_B$.
    \item $C1$ (or equivalently $C2$) performs local operations on $q^m_A$ based on the classical bits produced by the repeater in step~2.
\end{myenumlist}
End-to-end entanglement exploits \textit{quantum teleportation}, which can transfer a quantum state from one qubit to another with the only constraint that the state on the original qubit is destroyed, thus without violating the ``no-cloning theorem''.
%
The step~2 above can be repeated as many times as needed along a path.
For example, if the alternative (dashed) path $R2$--$R3$ in \rfig{basics-1} is used, the procedure  is the same except that both $R2$ and $R3$ have to perform entanglement swapping and provide the resulting classical qubits to $C1$ (or $C2$).
The behavior described is compatible with \addedb{so-called \textit{first generation} (1G)} quantum repeaters~\cite{munro_inside_2015} that do not implement any form of \ac{QEC}, which is still in its infancy.

\myssec{System model}{background:model}

%
We represent a quantum network as a directed graph, with each node representing a quantum computer or repeater, and each edge representing a physical link for quantum communication, as illustrated in the bottom part of \rfig{basics-1}.
The edges are characterized by two attributes: the \textit{capacity}, which is the number of EPR pairs that can be generated in the unit of time for end-to-end entanglement; and the \textit{generation fidelity}, which is a measure of the ``quality'' of the EPR pair upon generation.
The fidelity is a metric between 0 and 1 that summarizes in a quantitative manner how close the state of one qubit is to another, where 1 means perfection.

\added{\noindent\underline{\textbf{QoS definition}}. Quantum applications express their requirements in terms of:
\begin{myinlinelist}
    \item requested (end-to-end) capacity, in EPR pairs/s, and
    \item minimum fidelity.
\end{myinlinelist}}

In practice, all the processes involved in end-to-end entanglement introduce some undesirable effects\removed{: some of them are because of technology imperfections, that may be reduced or even eliminated in the future, while others depend on fundamental limitations, such as thermal noise and absorption/refraction phenomena}.
In this work we focus on two aspects that have a most significant impact on the performance of quantum networks with 1G repeaters, which results in a model widely adopted in the literature (e.g., \cite{chakraborty_entanglement_2020}).

First, the entanglement swapping operation performed by quantum repeaters is a stochastic process that succeeds with a given probability ($q$).
With current technology the failure probability is quite high, e.g., if linear optics components are used then $q$ is smaller than 50\%~\cite{sangouard_quantum_2011}.
\added{%
In a chain of hops, when one entanglement swapping fails, even at a single repeater, the overall end-to-end entanglement fails and all the intermediate EPR pairs are wasted.
}
\removed{
In other words, upon multiple attempts to create end-to-end entanglement from a source to the destination, the only one that succeeds is that where \textit{all} the entanglement swapping operations of the quantum repeaters along the path have succeeded;}
Thus, the success probability decreases exponentially with the number of intermediate hops.
Second, also the fidelity of the end-to-end entangled qubits decreases exponentially with the number of hops due to noise introduced by imperfect operations~\cite{briegel_quantum_1998}.
\removed{%
Overall, resource provisioning in quantum networks is different from that in classical networks.
In the latter, a traffic flow is characterized by a given requested rate (in bits/s) and, to determine whether it can be satisfied along a path, it is sufficient to check that there is enough residual capacity on each intermediate link; the amount of resources consumed, thus, increases linearly with the path length.
On the other hand, in quantum networks a traffic flow is characterized by a given requested capacity (in EPR pairs/s) and a minimum fidelity:
\begin{myinlinelist}
\item the actual capacity that will be used by the flow is inflated exponentially depending on the path length, and that capacity has to be available on \textit{each} intermediate link; and
\item the fidelity of the end-to-end entangled qubits is reduced exponentially with the path length, and that must remain above the minimum value specified.
\end{myinlinelist}
This has profound implications on \ac{QoS} and provisioning of resources, as well as on the overall performance of the network, which are studied in \rsec{contribution} after a brief survey of the literature in the next section.
}
\added{%
In \rsec{contribution} we analyze in quantitative terms the implications of the above features on \ac{QoS} through simulations in varied conditions.
Our goal is to raise awareness in the research community of the profound differences between quantum and classical networks, for what concerns not only  physical-layer technologies but also higher level aspects such as the provisioning of network resources.
}
  \section{State of the Art}%
  \label{sec:soa}%
  In the literature the problem of ``quantum routing'' has been well investigated.
It is commonly formulated as follows: find the best path to establish end-to-end entanglement between two nodes under given application constraints, usually the expected rate, in EPR pairs/s, and a minimum fidelity.
Many of the works have adapted known results from classical data networks to the quantum realm.
For instance, the famous Dijkstra's shortest-path algorithm was used in \cite{van_meter_path_2013}, which focused on the selection of the most suitable metric to use.
%
In this work we build on the findings in the literature on path selection in quantum networks.
In particular, as described in \rsec{cont:sim}, we use Dijkstra's algorithm iteratively to minimize the resources and maximize the fidelity.
%

A complementary problem is that of deciding the order in which to assign paths, under the assumption that a set of requests are to be processed at the same time.
In this area, we have proposed an iterative approach in \cite{cicconetti2021request} that attempts to balance an efficient use of resources with fair access to the network, while in \cite{chakraborty_entanglement_2020} the authors have proposed a split approach where the achievable rates for all traffic flows are determined first, and then rates are mapped to paths.
In this work we consider a pure online system, that is we assume that the requests of activating new traffic flows arrive one at a time and are served immediately\removed{ (no batch execution, no queuing, no reserve list, \ldots)}.
The reason for this is that we aim at elaborating on macroscopic aspects of resource provisioning, which have not yet been addressed in the literature.
As discussed in \rsec{conclusions}, future works will investigate more complex situations, possibly also considering the selection of the order in which to serve new flows in relation to the findings in the literature.%

  \section{Quality of Service in Quantum Networks}%
  \label{sec:contribution}%

In this section we introduce the simulation methodology (\rsec{cont:sim}), then delve into the analysis of the results \added{in two types of experiment, aimed at studying aspects related to admission control function (\rsec{cont:admission}) and network provisioning (\rsec{cont:provisioning})}.

\myssec{Simulation tool and methodology}{cont:sim}

%
We have developed a custom C++ simulator of the system model described in \rsec{background}, released as \textit{open source} on GitHub together with the scripts to fully reproduce the experiments \added{(repository \texttt{ccicconetti/quantum-routing}\footnote{\addedb{\url{https://github.com/ccicconetti/quantum-routing}}}, git tag \texttt{v1.4}, experiment label \texttt{003})}.
The tool simulates a dynamic system where new flows request admission following a Poisson distribution with a given arrival rate.
A traffic flow is characterized by the source and destination nodes and its \ac{QoS}, defined in terms of the requested rate ($r$, in EPR pairs/s) and the minimum end-to-end fidelity ($F$).
\removed{Based on calibration results (not reported here) we have chosen four classes of \ac{QoS} given by all the combinations of $r \in \{1, 10\}$ and $F \in \{0.7, 0.9\}$, which characterize traffic categories with well distinct features for the purposes of this work.}
\added{Four classes of \ac{QoS} are considered, as given by all the combinations of $r \in \{1, 10\}$ and $F \in \{0.7, 0.9\}$.}

\added{%
We assume that $q$ and $F_{init}$ are the same for all links, in particular it is $q=0.5$ and $F_{init}=0.95$ unless specified otherwise.
Under this assumption, the selection of the best path of an incoming traffic flow within the available residual capacity of the network can be greatly simplified.
In fact, selecting the shortest path from source to destination minimizes the gross rate and maximizes the fidelity at the same time.
The \textit{gross rate} is defined as the actual capacity that must be reserved to meet the minimum requested rate considering that entanglement swapping at intermediate nodes may fail.
Based on this observation, we have implemented the following admission control procedure in the simulator for a given new flow (which uses a temporary network graph of edge capacities):
}
%
%
\begin{myenumlist}
    \item Select the shortest path from source to destination\removed{, in terms of hops, which minimizes the resource utilization \textit{and} maximizes the fidelity of the end-to-end entangled pair; if there is no path, then the admission control fails}.

    \item Determine the \textit{gross rate} along the path selected\removed{, which is the number of EPR pairs/s that must be reserved to meet the minimum requested rate considering the entanglement swapping success probability ($q$) along the intermediate links}.

    \item 
    \added{The shortest path selected in the previous step may or may not have sufficient capacity to serve the requested rate $r$:
    \begin{myitemlist}
        \item if the capacity is sufficient, then go to next step~4;
        \item otherwise, remove the edge with minimum capacity from the temporary graph, to make sure that the same path is not selected again,and restart from step~1.
    \end{myitemlist}
    }

    \item Compute the fidelity of the end-to-end entangled pair:
    \begin{myitemlist}
        \item if it is greater or equal than the minimum requested value, then the flow can be admitted and all the capacities of the edges along the path are updated by removing the gross rate of the new flow;
        \item otherwise, the admission control fails.
    \end{myitemlist}
\end{myenumlist}
%
%
%
\added{The algorithm includes a loop between steps 1 and 3, which in the worst case may have to be executed a number of times equal to the number of edges.
For our purposes, its efficiency is irrelevant, as we are only concerned with the QoS-related results obtained to highlight the distinguishing features of quantum networks with respect to classical networks.}
A flow rejected is dropped permanently, while an admitted flow leaves the system after a duration drawn from a random variable exponentially distributed with mean \addedb{10~s}.

\begin{figure*}[tb]
    \centering
    \includegraphics[width=0.3\textwidth]{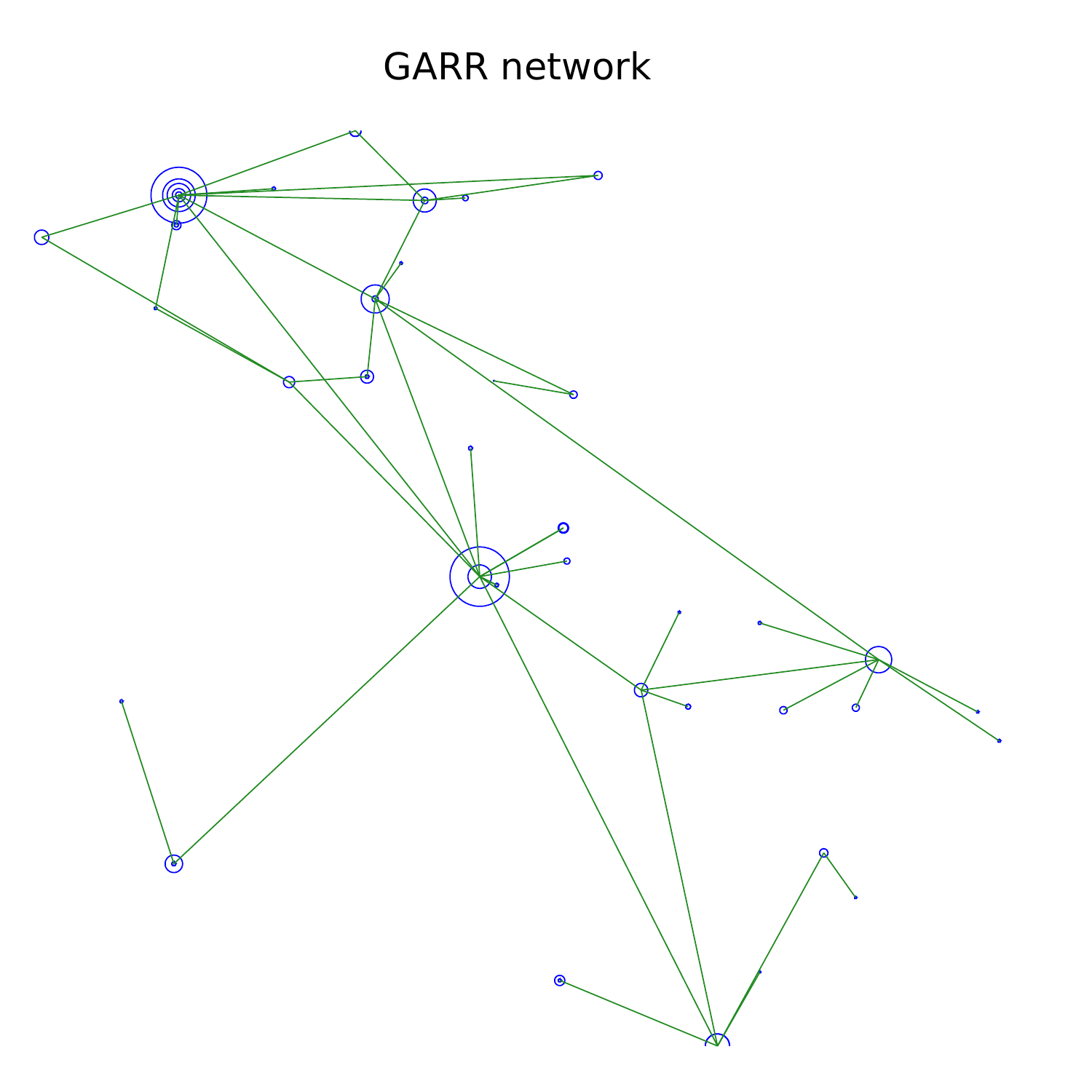}
    \includegraphics[width=0.3\textwidth]{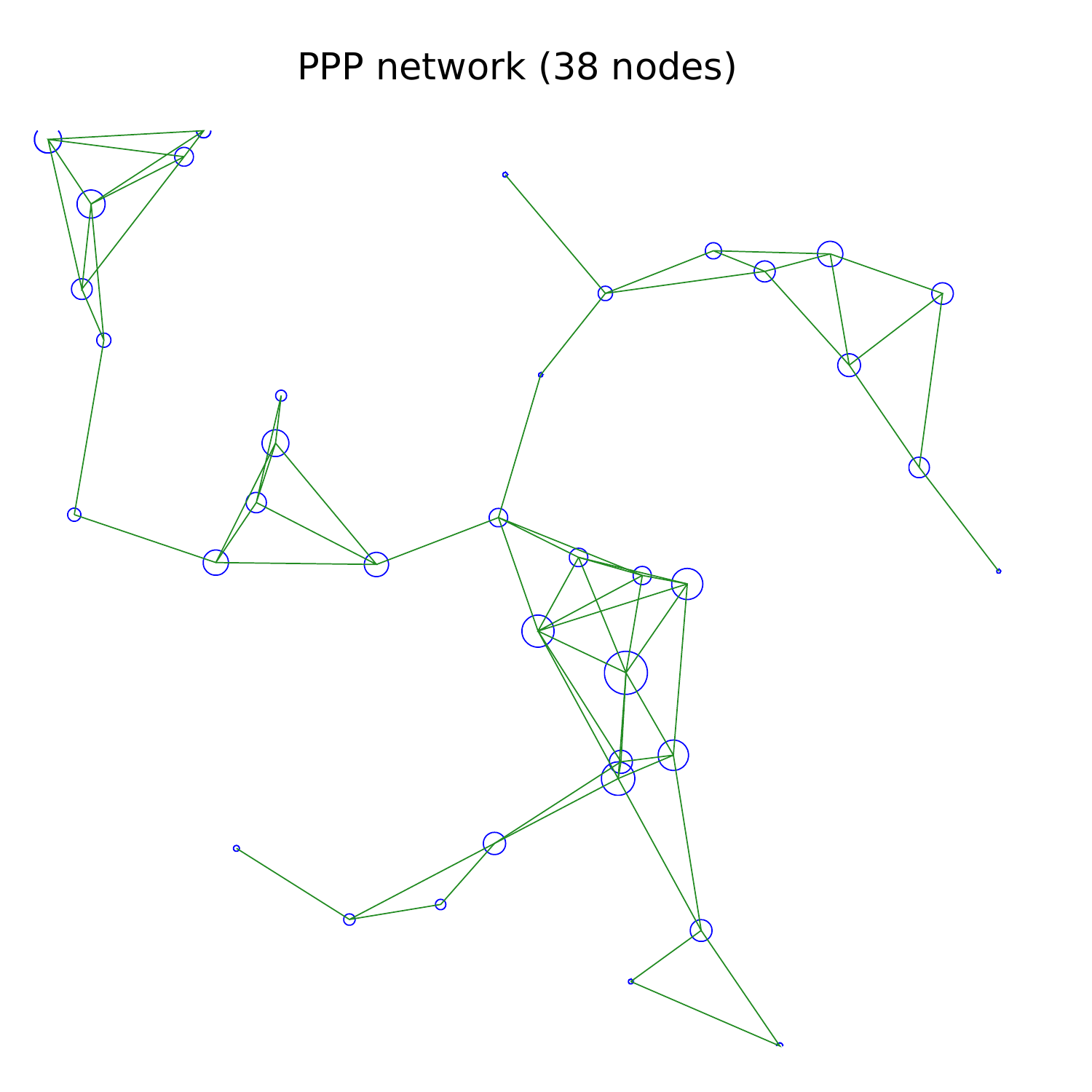}\\
    \includegraphics[width=0.3\textwidth]{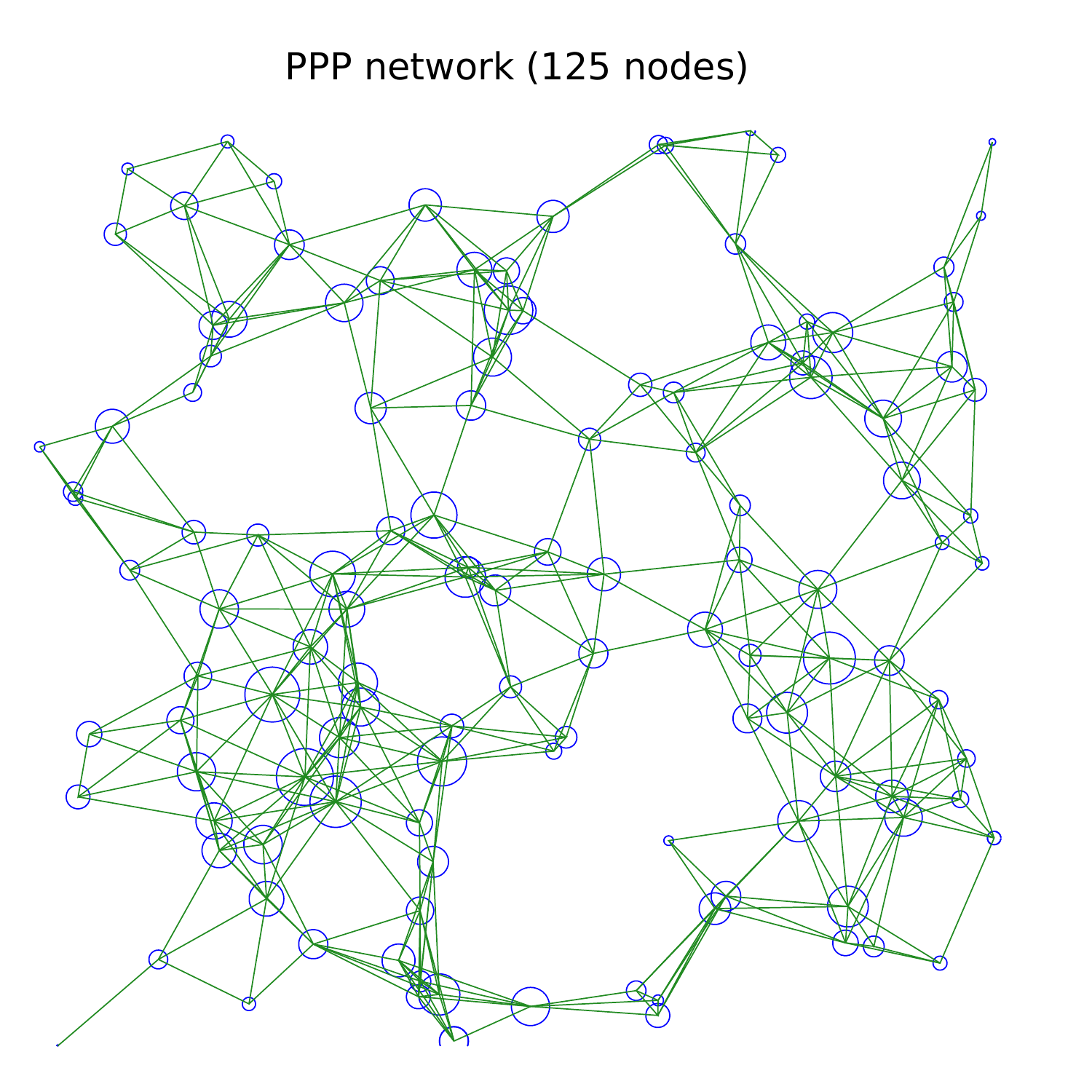}
    \includegraphics[width=0.3\textwidth]{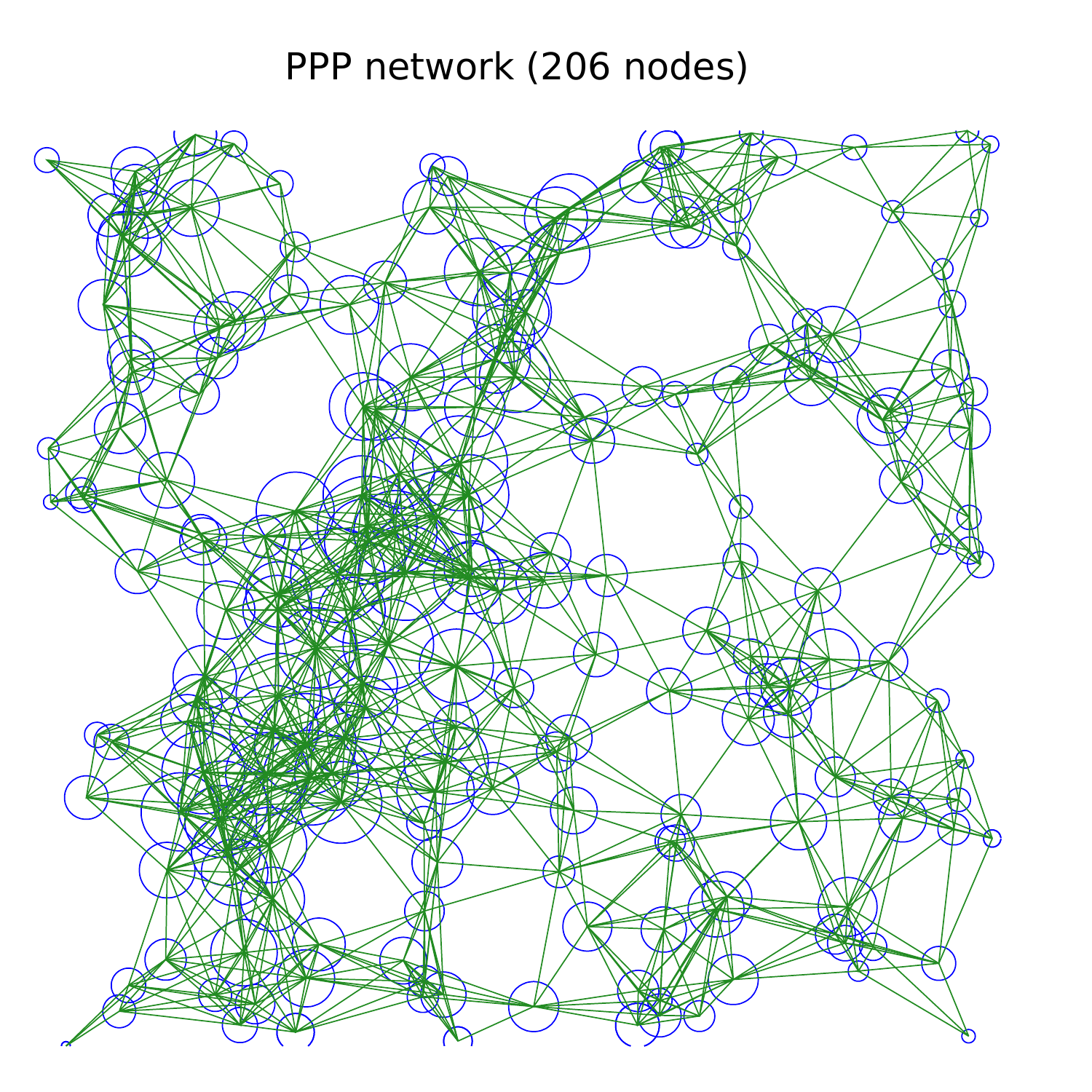}
    \caption{Examples of network topologies used:
    the GARR network in the first batch of experiments (top left);
    random PPP in the second batch with 38 (top right), 125 (bottom left), and 206 (bottom right) nodes.
    The radius of the circle around each node is proportional to the sum of the capacities of its links.}
    \label{fig:topo}
\end{figure*}

For the network topology we use two different models, which are described in the next section.
In both cases the capacity of a link is drawn from a random variable uniformly distributed between 1~EPR pair/s and a maximum number of EPR pairs/s, as in~\cite{chakraborty_entanglement_2020}.
\removed{%
Unless specified otherwise, the generation fidelity ($F_{init}$) is 0.95 and the probability that an entanglement swapping succeeds ($q$) is 0.5.
}
%
%
\added{%
The network topologies should not be intended as immediately matching today's (or tomorrow's) quantum networks, as we cannot envision the properties of the first operational quantum networks that will emerge based on the current status, still immature, of the technology of quantum links and repeaters.
Furthermore, we implicitly assume that fiber optic is used as carrier: however, our findings are more general since they consider fundamental aspects that, very likely, all 1G quantum networks will have in common.
}

\myssec{Admission control}{cont:admission}

\myfigeps{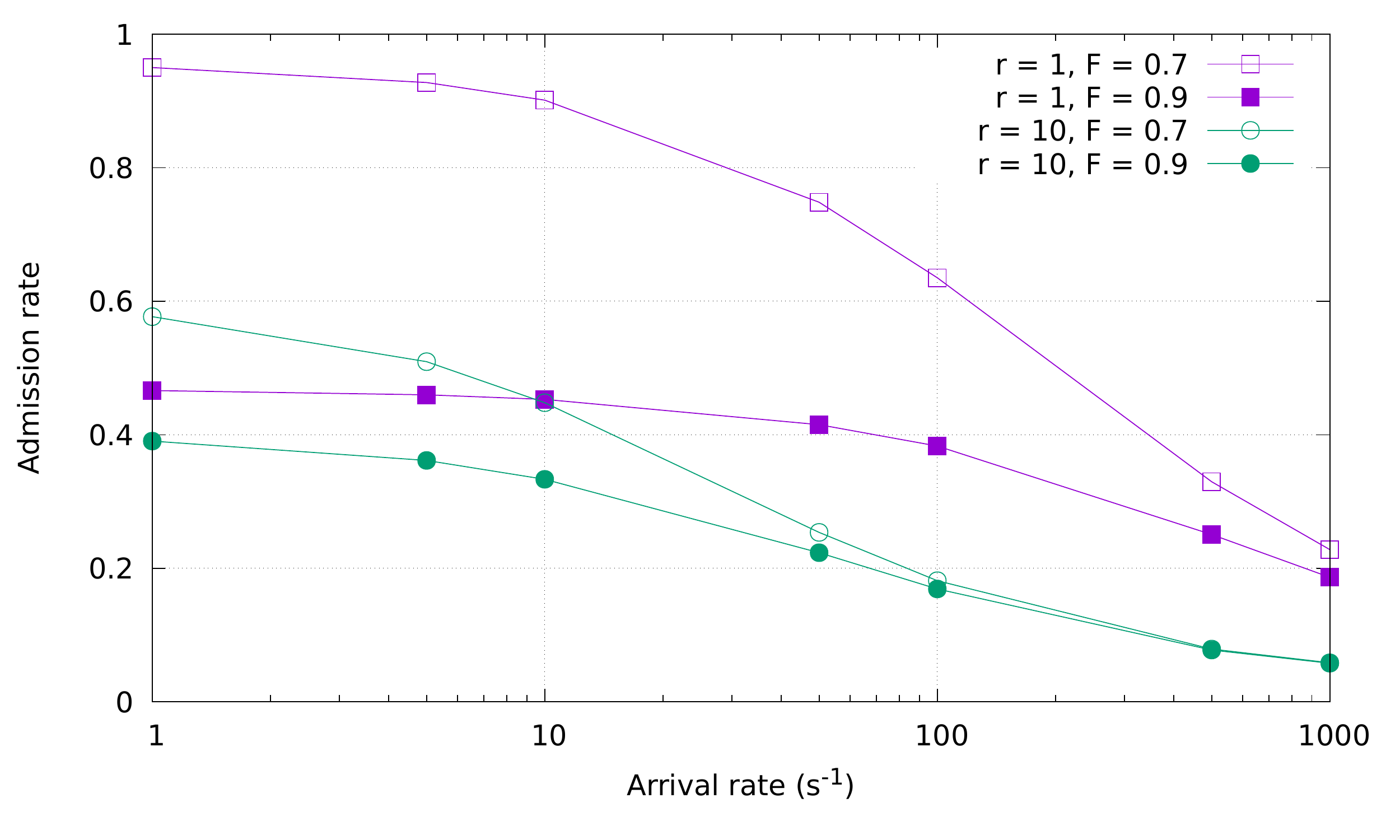}{GARR network: admission rate for the four classes of flows with increasing arrival rate (log scale).}

In the first set of experiments we use a real Internet topology, in particular the Italian national computer network for universities and research (GARR), which is illustrated as a graph in the top left part of \rfig{topo}, with nodes positioned in their geographical coordinates.
Each traffic flow is assigned randomly to one of the four classes of flows described in \rsec{cont:sim} and we consider an arrival rate increasing from 1~flow/s to 1,000~flows/s.
%
%
The source and destination nodes are selected randomly from the set of nodes in the network in a uniform manner.
In \rfig{003-load-admission-rate} we show the admission rate for the four \ac{QoS} classes.
For instance, admission rate 0.8 means that on average 80\% of the traffic flows of that class are admitted.
The results raise several points.

First, even at very low arrival rate (1 flow/s) the admission rate of all classes except $r=1,F=0.7$ is rather low, i.e., it is between 0.4 and 0.6, even though the total capacity in the network exceeds by far the individual rates requested.
\removed{%
This means that either the requested rate or the minimum fidelity indicated are often incompatible with what the network can offer.}
\added{This is because the gross rate grows exponentially (the fidelity decreases exponentially) with the path length, thus far away nodes require a high capacity even for modest requested rates (and may be unreachable because of fidelity constraints).}
%

Second, aggregated requests are penalized.
Consider for instance the two curves with $F=0.7$ and take the arrival rate of 10~flows/s for $r=1$ and 1~flow/s for $r=10$, which have the same total aggregated requested rate of 10~EPR pairs/s: the admission rate is much higher for $r=1$ (0.90) than it is for $r=10$ (0.57).
This remains true for all the values of $r$, and also by considering the two curves with $F=0.9$ (it is possible to visualize the transformation on the plot by imagining the green curves shifted \addedb{by 10~s} to the right on the $x$-axis).
This is because in the network there might be several paths with similar length connecting the two nodes wishing to establish end-to-end entanglement, which can all be used by multiple small requests but not all together by the same large traffic flow.
Thus, the use of \textit{multipath}, which in classical data networks is typically discouraged as it leads to increased complexity and out-of-order delivery of packets, can be revamped in quantum networks to reduce the penalty of large traffic flows compared to small ones, as mentioned in~\cite{pant_routing_2019}.

\myfigeps{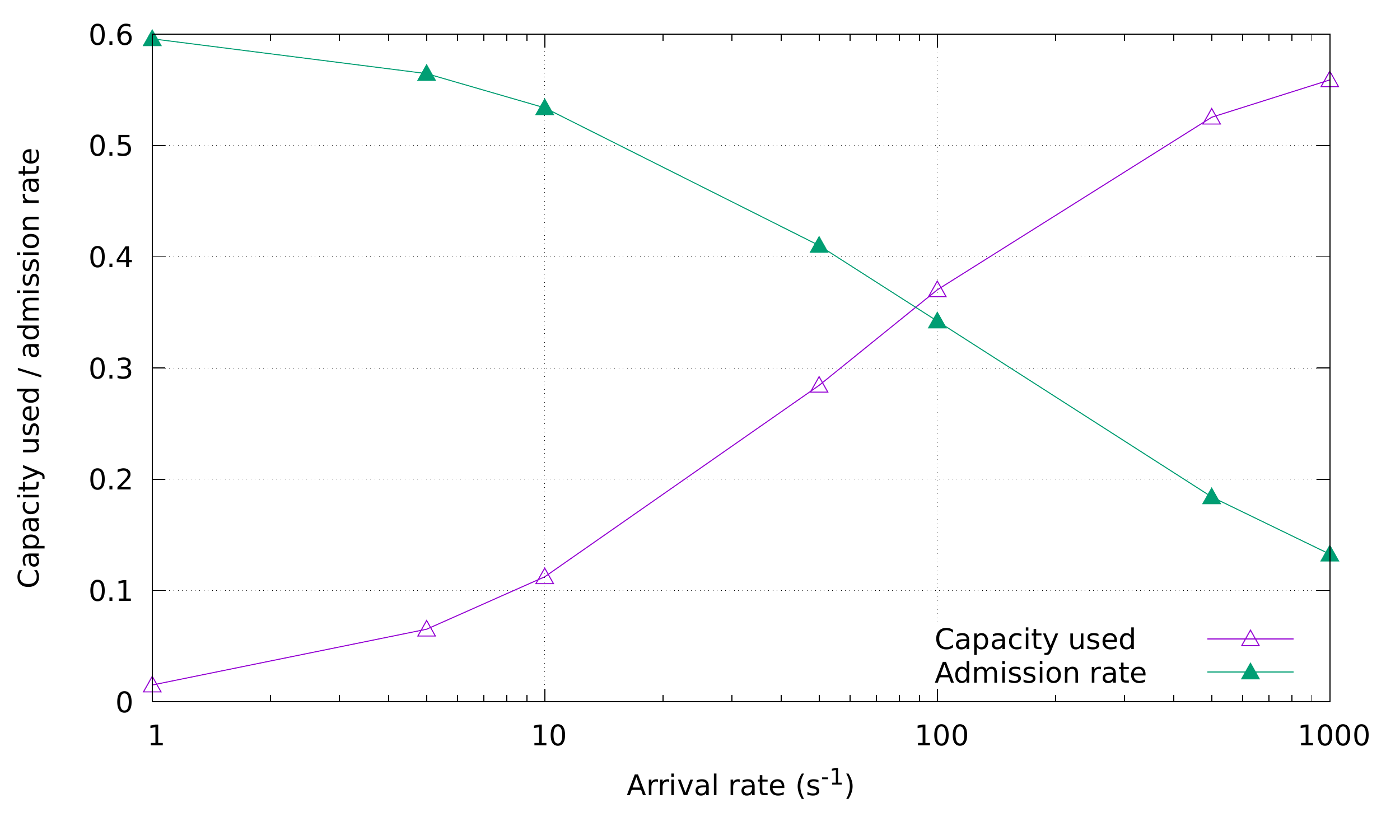}{%
GARR network:
capacity used vs.\ total admission rate with increasing arrival rate (log scale).}

Third, for all the traffic flows, including $r=1,F=0.7$, the admission rate decreases exponentially with the arrival rate; note that the $x$-axis in \rfig{003-load-admission-rate} is in logarithmic scale.
In other words, as an increasing number of traffic flows become active in the network, it becomes more and more difficult to admit new flows.
We speculate that this non-linearity would make it difficult in practice to provision the resources, e.g., to estimate how many traffic flows can be served under a given target admission rate: small deviations from the planned use of resources may have a dramatic impact on the service offered.

Even worse is the fact that this happens when the overall network resources are largely underutilized.
Consider \rfig{003-load-capacity-vs-admission}, which plots the average admission rate over all the \ac{QoS} classes vs.\ the capacity used, as the load increases.
The \textit{capacity used} is defined as the average amount of resources reserved by active traffic flows divided by the total network capacity.
As can be seen, the admission rate drops steeply already when a small fraction of the available capacity is used: with a mere 10\% of the capacity used, the admission rate is already about 50\%, and even with the maximum load considered, when the admission rate drops to a little more than 10\%, the capacity used is just 56\%.
This is in contrast with what happens in classical data networks, where the admission rate is usually a \textit{smooth} function of the load under nominal conditions.

\added{%
\noindent\underline{\textbf{Takeaway messages.}}

\begin{myitemlist}
    \item The exponential increase of the gross rate / exponential decrease of the fidelity with the path length practically limit the distance between communicating parties in a quantum network: in the near future, dense small networks will be less preferable than large sparse ones, possibly using satellite links to extend coverage.
    \item Aggregated requests (\textit{elephants}) are penalized compared to smaller requests (\textit{mice}): multipath can be investigated to address this issue.
    \item Admission control is more erratic than in classical networks: operators should consider the non-linear/steep function response and devise service plans accordingly.
\end{myitemlist}
}

\myssec{Network provisioning}{cont:provisioning}

We now focus on quantum network provisioning.
To this aim, we use artificial topologies, like in~\cite{dai_optimal_2020}, generated via a \ac{PPP} by dropping a number of nodes, with given average, in a flat square grid with edge size 100~km. A link is then added between two nodes if their Euclidean distance is smaller than 15~km, which is a realistic value based on current quantum communication network technology~\cite{yu_entanglement_2020}.
Examples of the resulting topologies are illustrated in \rfig{topo} (top right and bottom).
We consider uniform traffic compositions, with a single \ac{QoS} class.
The source and destination pairs are assigned in one of the following ways:
\begin{myinlinelist}
    \item \textit{uniform}: the source and destination pairs are selected randomly from the set of nodes in the network in a uniform manner, like in the first set of experiments;
    \item \textit{weighted}: each node has a probability to be a source (or a destination) of a traffic flow that is proportional to the sum of the capacities of the links connected to that node.
\end{myinlinelist}
The weighted assignment is used to mimic a system where the network infrastructure was provisioned based on an estimate of the future (average) traffic demands.

\myfigfulleps[width=\textwidth]{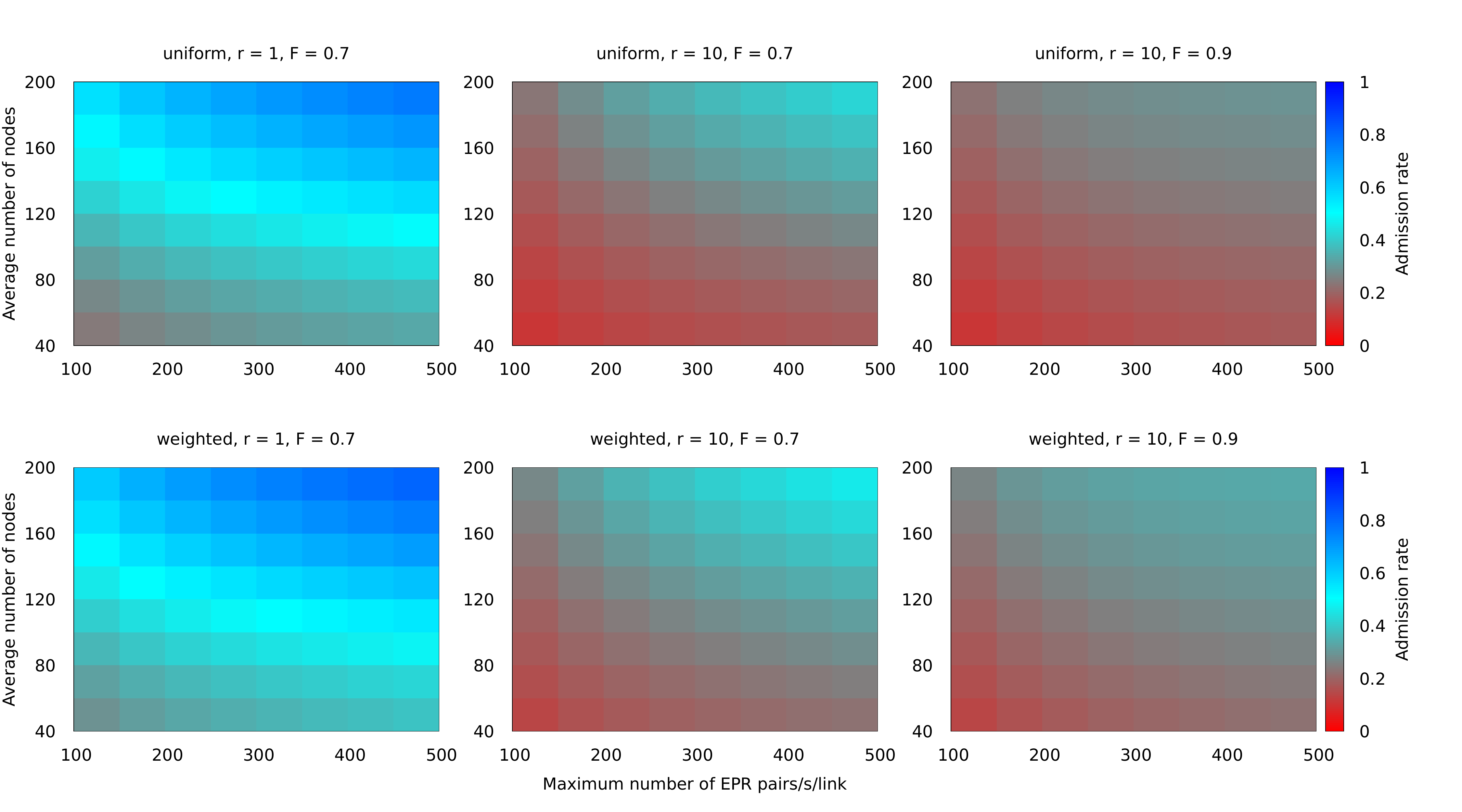}{%
PPP network:
admission rate for the three classes of flows (left: $r=1,F=0.7$; center: $r=10,F=0.7$; right: $r=10,F=0.9$), with increasing resources provisioned ($x$-axis: maximum number of EPR pairs/s/link, $y$-axis: average number of nodes), when selecting uniformly the source/destination pairs (top) vs.\ preferring nodes with a higher aggregate capacity of their links (bottom); the arrival rate is 100 flows/s.}

In \rfig{003-net-admission-rate} we show the admission rate of three \ac{QoS} classes (left: $r=1,F=0.7$; center: $r=10,F=0.7$; right: $r=10,F=0.9$) for the two assignment strategies.
The arrival rate is, on average, 100 flows/s, with increasing link capacity ($x$-axis) and number of nodes ($y$-axis).
In all the plots an increasing $x$-axis or $y$-axis value corresponds to a higher density of network infrastructures, hence higher deployment and operational costs.
On the other hand, the color maps to the admission rate from red (low) to blue (high), which in turn significantly affects the customer experience.

The admission rate severely degrades as the application \ac{QoS} requirements become more demanding, both requested rate (from $r=1$ to $r=10$) and fidelity (from $F=0.7$ to $F=0.9$).
Indeed, reasonable admission rates (around 50\%) can only be achieved with many nodes and a high average link capacity, regardless of whether a uniform or weighted assignment is used, where the latter only slightly improves the performance.
With more relaxed application \ac{QoS} requirements, i.e., $r=1,F=0.7$ (left plots in \rfig{003-net-admission-rate}), the admission rate is significantly better, especially with a weighted assignment policy.
It is interesting to note that this metric improves regularly at each step in the two axes.
In other words, from the point of view of the applications, it is the same if the link capacity is increased by 50 EPR~pairs/s ($x$-axis) or 20 nodes are added to the network ($y$-axis).
However, the provisioning cost along the two directions can be very different for the quantum network operator:

\begin{myitemlist}
    \item Adding a \textit{new node} means that a site equipped with quantum repeaters is added to the network.
    In the simplest case, this can be as straightforward as adding a new rack-mountable piece of equipment to a data center or \ac{PoP} that is already interconnected to a fiber optic infrastructure used for quantum and classical communications.
    However, the transmission of qubits is in general more fragile than that of classical information.
    Thus, we can also expect that quantum repeaters will be needed in new sites to be provisioned specifically for quantum communication, which would hugely increase the per-node deployment cost.
    \item The \textit{link capacity} depends on the amount of physical carriers used, e.g., fiber optic cables, and on the number of devices used to generate EPR pairs and to entangle flying and matter qubits (see \rsec{background:basics}), most likely involving single photon detection.
    This suggests that deployment and operation costs should scale linearly with link capacity, but the statement will have to be confirmed when mass production of 1G quantum repeaters begins.
\end{myitemlist}





\myfigeps{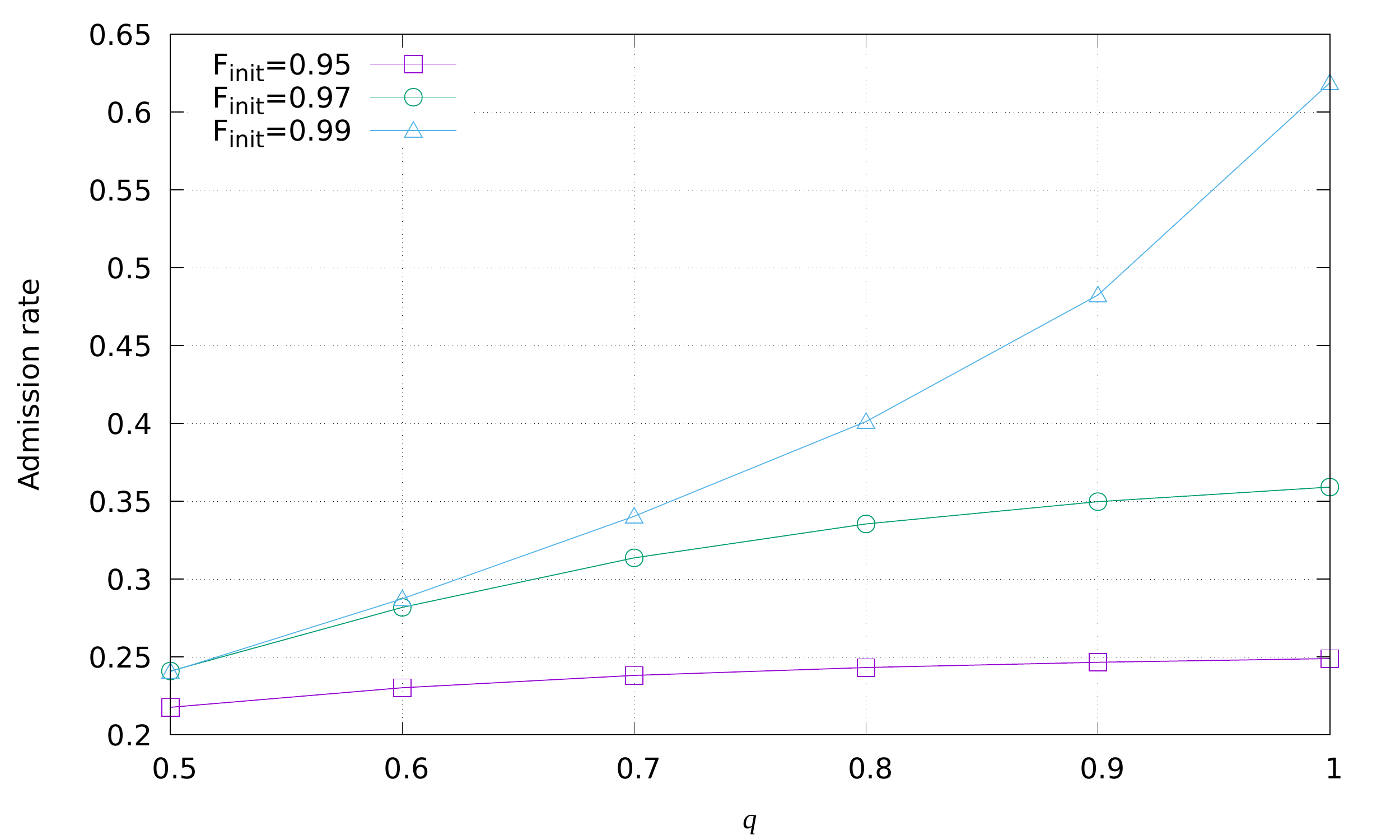}{PPP network: admission rate with 120 nodes (on average) and 300 maximum EPR/s/link, with increasing entanglement swapping success probability ($q$) and different values of generation fidelity ($F_{init}$); the arrival rate is 100 flows/s, with $r=10$ and $F=0.9$.}

To complete our analysis, in \rfig{003-gen-admission-rate} we show the admission rate obtained when considering quantum repeaters that \textit{exceed} the technology limitations expected for near-term devices, i.e., with entanglement swapping success probability ($q$) increasing from 0.5 to 1 and fidelity of freshly generated EPR pairs $F_{init} \in \{ 0.95, 0.97, 0.99\}$.
We do so in a PPP network with the same characteristics as in the previous scenario, but with a fixed maximum link capacity (300 EPR~pairs/s) and average number of nodes (120); furthermore, we only consider traffic flows with $r=10,F=0.9$, which are the most challenging \ac{QoS} requirements.
The plot shows that $F_{init}=0.95$ hampers the admission rate for all values of the entanglement swapping success probability.
Already with $F_{init}=0.97$ the admission rate is allowed to grow from 25\% to about 35\%, but still with a sub-linear trend.
Instead, with $F_{init}=0.99$ the admission rate is not constrained noticeably by the minimum fidelity requested $F$, hence it is limited only by $q$.
In particular, when $q=1$, the provisioning process becomes the same as with classical communications, and hence can be considered an upper bound of the achievable performance.
%

\added{%
\noindent\underline{\textbf{Takeaway message.}}
The customer experience, in terms of the blocking probability, depends significantly on the network resources provisioned.
The latter can be expanded along two directions, with similar effect from the users' perspective: adding new nodes or increasing the capacity of existing links, which likely will have different costs and practical implications.
The evolution of quantum repeaters will soften the network provisioning difficulties, but only in the long-term, i.e., when the entanglement swapping success probability and generation fidelity will be close to 1.
}%


%
  \section{Summary and Open Issues}%
  \label{sec:conclusions}%
  In this paper we have investigated some fundamental aspects of \ac{QoS} and provisioning in quantum networks with 1G repeaters.
Our work extends the state of the art on quantum routing by providing a new perspective from the point of view of the network operator, which in the future will have to cope with the definition of suitable business models and service plans based on the estimated revenues and costs.
\removed{The results have shown that even when the network is lightly loaded there is an unexpectedly high fraction of traffic flows that cannot be served, because their \ac{QoS} constraints are met, due to the exponential growth of the network resource demands and the exponential decrease of the fidelity with the path length.}
\added{Because of the fragility of qubits and the impossibility to copy them, network resource demands increase exponentially with the path length, while the fidelity decreases exponentially.
Under these conditions, simulation results have shown that, even when the network is lightly loaded, there is an unexpectedly high fraction of traffic flows that cannot be served because their \ac{QoS} constraints are not met.}
This becomes even worse at high network loads, where the unused capacity is difficult to reach, both with completely random source/destinations and when nodes with greater capacity are more likely to originate traffic flows.
\removed{Such a phenomenon leads to potential unfairness between traffic flows with different \ac{QoS} requirements, and in general it suggests that quantum network resources have to be greatly overprovisioned, which can dramatically hamper the deployment of the Quantum Internet.}
The evolution of technology, e.g., to increase the generation fidelity and reduce the probability of failed entanglement swapping in quantum repeaters, will alleviate the problem, but this cannot be expected to happen soon.

In our work, we have unveiled only the tip of a new iceberg of research with many questions left unanswered:
\begin{myitemlist}
    \item What is the impact of noise models (e.g., for decoherence of flying qubits over distance) on top of the effects considered in this paper?
    \item Distillation is a procedure that increases the fidelity of end-to-end qubits, at the cost of reducing the capacity further: is this an application-level feature or should the network provide distillation as a service?
    \item Based on the results obtained, multipath may help to better use the network resources, but what is the downside in terms of control/management complexity?
    \item We have considered a pure online model, where each new traffic flow is served immediately or dropped, but what if the quantum network operator wished to pre-reserve paths based on expected traffic demands?
    \item Does the definition of \ac{QoS} capture well the requirements of real applications and what are their associated business models?
    \item Some applications, e.g., distributed \ac{QC}, may require a variable rate of EPR pairs: what is the impact in terms of provisioning and can different types of application co-exist peacefully in the same network?
    \item \added{What is the role of time synchronization in the network?}
    \item \added{In principle, EPR pairs can be prepared and stored in quantum memories for later use: how to best use such \textit{reservoirs} of resources that have no classical counterpart?}
\end{myitemlist}

\added{To conclude, inherent properties of quantum networks may greatly affect \ac{QoS} requirements and provisioning.
In particular, the potential unfairness between traffic flows with different characteristics has to be addressed, and in general the operators may have to cope with the need of greatly overprovisioning resources.
If not addressed properly, these features can dramatically hamper the deployment of the Quantum Internet.}

\section*{Acknowledgment}

This work was co-funded by European Union, \textit{PON Ricerca e Innovazione 2014-2020 FESR/FSC}, project ARS01\_00734 QUANCOM.




\begin{acronym}
  \acro{3GPP}{Third Generation Partnership Project}
  \acro{5G-PPP}{5G Public Private Partnership}
  \acro{AA}{Authentication and Authorization}
  \acro{ADF}{Azure Durable Function}
  \acro{AI}{Artificial Intelligence}
  \acro{API}{Application Programming Interface}
  \acro{AP}{Access Point}
  \acro{AR}{Augmented Reality}
  \acro{BGP}{Border Gateway Protocol}
  \acro{BSP}{Bulk Synchronous Parallel}
  \acro{BS}{Base Station}
  \acro{CDF}{Cumulative Distribution Function}
  \acro{CFS}{Customer Facing Service}
  \acro{CPU}{Central Processing Unit}
  \acro{DAG}{Directed Acyclic Graph}
  \acro{DHT}{Distributed Hash Table}
  \acro{DNS}{Domain Name System}
  \acro{ETSI}{European Telecommunications Standards Institute}
  \acro{FCFS}{First Come First Serve}
  \acro{FSM}{Finite State Machine}
  \acro{FaaS}{Function as a Service}
  \acro{GPU}{Graphics Processing Unit}
  \acro{HTML}{HyperText Markup Language}
  \acro{HTTP}{Hyper-Text Transfer Protocol}
  \acro{ICN}{Information-Centric Networking}
  \acro{IETF}{Internet Engineering Task Force}
  \acro{IIoT}{Industrial Internet of Things}
  \acro{ILP}{Integer Linear Programming}
  \acro{IPP}{Interrupted Poisson Process}
  \acro{IP}{Internet Protocol}
  \acro{IRTF}{Internet Research Task Force}
  \acro{ISG}{Industry Specification Group}
  \acro{ITS}{Intelligent Transportation System}
  \acro{ITU}{International Telecommunication Union}
  \acro{IT}{Information Technology}
  \acro{IaaS}{Infrastructure as a Service}
  \acro{IoT}{Internet of Things}
  \acro{JSON}{JavaScript Object Notation}
  \acro{K8s}{Kubernetes}
  \acro{KVS}{Key-Value Store}
  \acro{LCM}{Life Cycle Management}
  \acro{LL}{Link Layer}
  \acro{LOCC}{Local Operations and Classical Communication}
  \acro{LTE}{Long Term Evolution}
  \acro{MAC}{Medium Access Layer}
  \acro{MBWA}{Mobile Broadband Wireless Access}
  \acro{MCC}{Mobile Cloud Computing}
  \acro{MEC}{Multi-access Edge Computing}
  \acro{MEH}{Mobile Edge Host}
  \acro{MEPM}{Mobile Edge Platform Manager}
  \acro{MEP}{Mobile Edge Platform}
  \acro{ME}{Mobile Edge}
  \acro{ML}{Machine Learning}
  \acro{MNO}{Mobile Network Operator}
  \acro{NAT}{Network Address Translation}
  \acro{NISQ}{Noisy Intermediate-Scale Quantum}
  \acro{NFV}{Network Function Virtualization}
  \acro{NFaaS}{Named Function as a Service}
  \acro{NV}{Nitrogen-Vacancy}
  \acro{OSPF}{Open Shortest Path First}
  \acro{OSS}{Operations Support System}
  \acro{OS}{Operating System}
  \acro{OWC}{OpenWhisk Controller}
  \acro{PMF}{Probability Mass Function}
  \acro{PoP}{Point of Presence}
  \acro{PPP}{Poisson Point Process}
  \acro{PU}{Processing Unit}
  \acro{PaaS}{Platform as a Service}
  \acro{PoA}{Point of Attachment}
  \acro{PPP}{Poisson Point Process}
  \acro{QC}{Quantum Computing}
  \acro{QEC}{Quantum Error Correction}
  \acro{QKD}{Quantum Key Distribution}
  \acro{QIRG}{Quantum Internet Research Group}
  \acro{QoE}{Quality of Experience}
  \acro{QoS}{Quality of Service}
  \acro{RPC}{Remote Procedure Call}
  \acro{RR}{Round Robin}
  \acro{RSU}{Road Side Unit}
  \acro{SBC}{Single-Board Computer}
  \acro{SDK}{Software Development Kit}
  \acro{SDN}{Software Defined Networking}
  \acro{SJF}{Shortest Job First}
  \acro{SLA}{Service Level Agreement}
  \acro{SMP}{Symmetric Multiprocessing}
  \acro{SoC}{System on Chip}
  \acro{SLA}{Service Level Agreement}
  \acro{SRPT}{Shortest Remaining Processing Time}
  \acro{SPT}{Shortest Processing Time}
  \acro{STL}{Standard Template Library}
  \acro{SaaS}{Software as a Service}
  \acro{TCP}{Transmission Control Protocol}
  \acro{TSN}{Time-Sensitive Networking}
  \acro{UDP}{User Datagram Protocol}
  \acro{UE}{User Equipment}
  \acro{URI}{Uniform Resource Identifier}
  \acro{URL}{Uniform Resource Locator}
  \acro{UT}{User Terminal}
  \acro{VANET}{Vehicular Ad-hoc Network}
  \acro{VIM}{Virtual Infrastructure Manager}
  \acro{VR}{Virtual Reality}
  \acro{VM}{Virtual Machine}
  \acro{VNF}{Virtual Network Function}
  \acro{WLAN}{Wireless Local Area Network}
  \acro{WMN}{Wireless Mesh Network}
  \acro{WRR}{Weighted Round Robin}
  \acro{YAML}{YAML Ain't Markup Language}
\end{acronym}

\end{document}